\pdfoutput=1
\documentclass[9pt]{article}
\usepackage{ACL2023}
\usepackage{times}
\usepackage{graphicx}
\usepackage{booktabs}
\usepackage{arydshln}  
\newcommand{\Comment}[1]{}

\newcommand{\bi}{\begin{list}{$\bullet$}{
    \setlength{\leftmargin}{1.5 em}
    \setlength{\itemsep}{0 pt}
    \setlength{\topsep}{3 pt}
    \setlength{\parsep}{3 pt}
    \setlength{\partopsep}{0 pt}
    \setlength{\labelwidth}{1 em}
    \setlength{\labelsep}{0.5 em}
    \setlength{\parskip}{0cm}  }}
\newcommand{\ei}{\end{list}}

\newcommand{\BE}{\begin{enumerate}}
\newcommand{\EE}{\end{enumerate}}

\newcommand{\initab}{                           
\begin{tabbing}
XXX \= XXXX \= \kill
}
\newcommand{\begpub}{
\begin{quotation}
\noindent
}

\newcommand{\finpub}{
\end{quotation}
}

\newcommand{\shorten}[1]{} 

\newcommand\dan[1]{{#1}}

\newcommand\jb[1]{{#1}}

\newcounter{sect}
\newcounter{subsect}[sect]

\title{The Semantic Scholar Open Data Platform}

\author{
Rodney Kinney,
Chloe Anastasiades,
Russell Authur\thanks{\hspace{8pt}Work done while at the Allen Institute for AI} ,
Iz Beltagy\footnotemark[1],
Jonathan Bragg, \AND
Alexandra Buraczynski, 
Isabel Cachola\footnotemark[1] , 
Stefan Candra, 
Yoganand Chandrasekhar\footnotemark[1], \AND
Arman Cohan\footnotemark[1] , 
Miles Crawford\footnotemark[1] ,
Doug Downey,
Jason Dunkelberger,
Oren Etzioni\footnotemark[1], \AND
Rob Evans,
Sergey Feldman,
Christopher Fiorelli\footnotemark[1],
Daniel King\footnotemark[1],
Joseph Gorney\footnotemark[1], \AND
David Graham,
Dany Haddad,
Fangzhou Hu, 
Regan Huff, 
Daniel King\footnotemark[1] , \AND
Sebastian Kohlmeier\footnotemark[1] ,
Bailey Kuehl, 
Michael Langan,
Daniel Lin, 
Haokun Liu\footnotemark[1] , \AND
Kyle Lo, 
Jaron Lochner, 
Kelsey MacMillan,
Tyler Murray,
Chris Newell, \AND 
Smita Rao, 
Shaurya Rohatgi\footnotemark[1] ,
Paul Sayre, 
Zejiang Shen\footnotemark[1] ,
Amanpreet Singh, \AND
Luca Soldaini, 
Shivashankar Subramanian\footnotemark[1] ,
Amber Tanaka,
Alex D. Wade\footnotemark[1] , \AND
Linda Wagner\footnotemark[1] , 
Lucy Lu Wang\footnotemark[1] , 
Chris Wilhelm, 
Caroline Wu,
Jiangjiang Yang, \AND
Angele Zamarron, 
Madeleine Van Zuylen\footnotemark[1] ,
Daniel S. Weld \\
Allen Institute for Artificial Intelligence \\
Seattle, Washington \\
\texttt{\{rodneyk,dougd,danw\}@allenai.org} 
}

\begin{document}

\maketitle

\begin{abstract}
The volume of scientific output is creating an urgent need for automated tools to help scientists keep up with developments in their field. Semantic Scholar~(S2) is an open data platform and website aimed at accelerating science by helping scholars discover and understand scientific literature.  We combine public and proprietary data sources using state-of-the-art techniques for scholarly PDF content extraction and automatic knowledge graph construction to build the Semantic Scholar Academic Graph, the largest open scientific literature graph to-date, with 225M+ papers, 100M+ authors, 650M+ paper-authorship edges, and 2.8B+ citation edges. The graph includes advanced semantic features such as structurally parsed text, natural language summaries, and vector embeddings.  In this paper, we describe the components of the S2 data processing pipeline and the associated APIs offered by the platform. We will periodically update this document to reflect improvements and new data offerings.
\end{abstract}

\section{Introduction}

Semantic Scholar\footnote{https://semanticscholar.org} (S2),  was launched in 2015 by the Allen Institute for Artificial Intelligence\footnote{https://allenai.org} (AI2) to help scholars combat information overload and more efficiently discover and understand the most relevant research literature. Through a growing number of partnerships with scientific publishers and preprint services, Semantic Scholar has built a comprehensive and open corpus of scientific publications as a public service. While the Semantic Scholar website provides many features, such as automatically-generated author pages, personalized libraries, and paper recommendations, most of the site's data and functionality is also available via data download, open-source libraries, and API services. 

In this paper, we overview the primary technology used to build the corpus, and the API services and downloads we provide to access it. We hope that the resources described in this article can further accelerate a variety of work that depends critically on high-quality scholarly data.  Research into scientific NLP and science of science benefit directly from such resources. More generally, we are enabling the development of new applications that help scientists discover and understand the literature of their field.  The need for timely and comprehensive scholarly data has become more imperative since the 2021 sunsetting of the Microsoft Academic Graph (MAG) \cite{Sinha2015AnOO}, which was long a standard source for scholarly data in the community. 

The remainder of the paper proceeds as follows.
In Section \ref{sec:platform}, we give an overview of the data and services that comprise the platform. In Section \ref{sec:pipeline}, we describe the platform's data processing pipeline. In Section \ref{sec:online_resources}, we describe our public APIs and downloadable datasets.  In Section \ref{sec:related}, we discuss related work. In Section \ref{sec:conclusion}, we  conclude and discuss future work.

\begin{figure*}
\begin{centering}
\includegraphics[width=\linewidth]{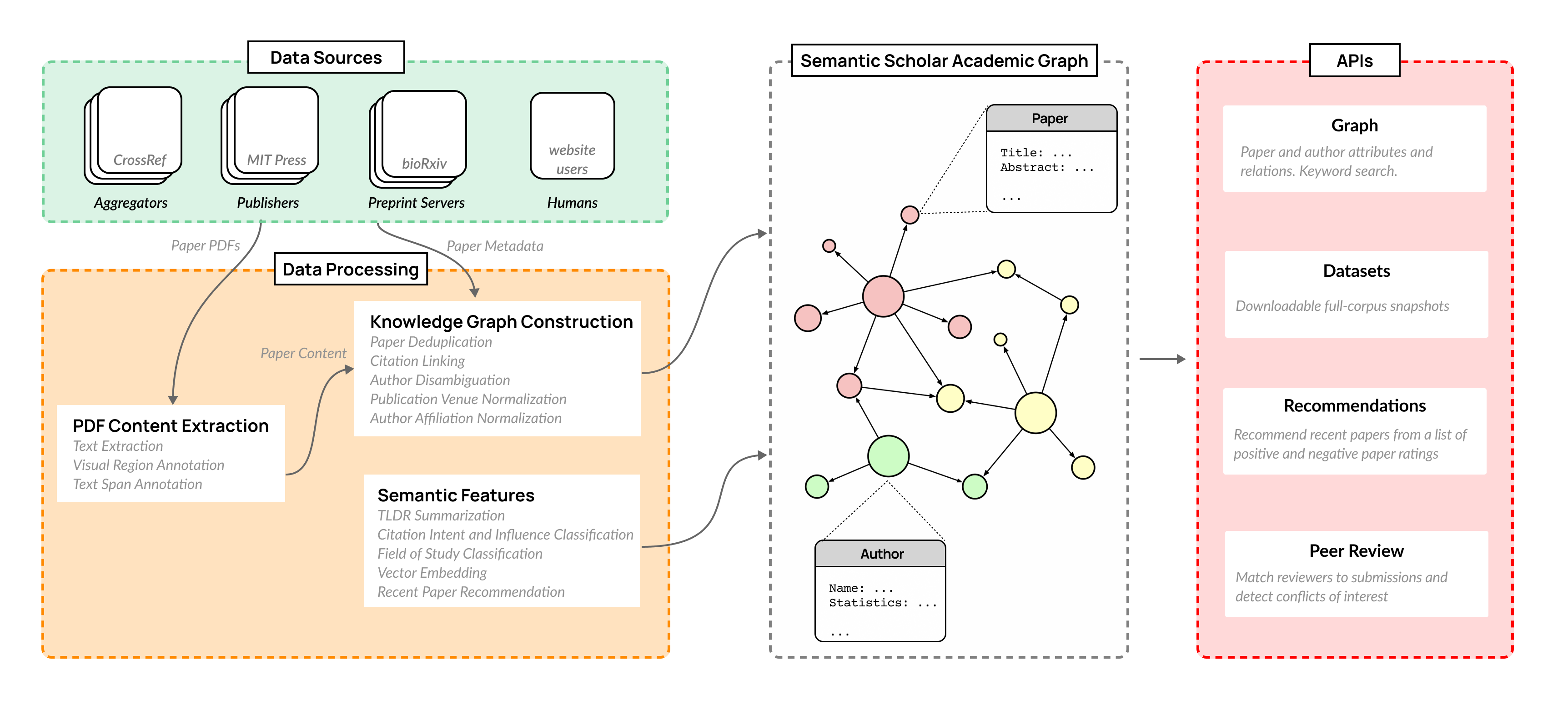}
\end{centering}
\vspace*{-0.15in}
\caption{Illustration of the Semantic Scholar platform }
\label{fig:platform}
\end{figure*}

\section{Platform Overview}
\label{sec:platform}
The purpose of the Semantic Scholar Open Data Platform is to build and distribute the Semantic Scholar Academic Graph, or S2AG (pronounced "stag"). S2AG is a disambiguated, high-quality, bibliographic knowledge graph. The nodes of S2AG represent papers, authors, venues, and academic institutions. The edges represent papers written by an author, papers cited by another paper, papers published in a venue, and authors affiliated with an institution. S2AG is built by ingesting a variety of data sources into our data processing pipeline, and is distributed via a variety of publicly-available APIs and datasets, as illustrated in Figure \ref{fig:platform}.

The work we describe here is an update of a previous version of the Semantic Scholar data pipeline described in \citet{Ammar2018ConstructionOT}.  The core structure of the graph (authors, papers and their relationships) is the same, although the components that build the knowledge graph have been refined. The entity extraction and linking described in \citet{Ammar2018ConstructionOT} have been deprecated in favor of the semantic features described below.

Table~\ref{tab:freq} summarizes the overall size of S2AG as of January 2023.\footnote{We are populating the author-affiliation links at the time of writing, so we have given the expected eventual number.} S2AG covers many natural, physical, and social sciences. Table \ref{tab:fos} summarizes the number of paper records in different academic fields of study. Some papers are assigned to multiple fields, whereas many are unclassified (``n/a''), due to lack of sufficient information.

\section{Data Processing Pipeline}
\label{sec:pipeline}

At the core of Semantic Scholar lies a sophisticated data processing pipeline that continually ingests documents and metadata from numerous sources, extracting full text and metadata from PDFs, normalizing and disambiguating authors, institutions, and venues, classifying each paper's field of study, generating a textual summary of its key results, and more. 

The pipeline builds S2AG by ingesting academic paper metadata and PDF content from a variety of \emph{Data Sources}. A \emph{PDF Content Extraction} system extracts structured data from unstructured PDFs. The extracted content, along with structured metadata from the input sources, is processed by a series of \emph{Knowledge Graph Construction} systems that build S2AG. A set of models add \emph{Semantic Features} to the graph, such as such as paper summarization and vector embeddings.

Many components of our pipeline are available as open software or models. Table~\ref{tab:models} provides a high-level summary alongside links to publicly available code, models and datasets, where applicable. Many of these also have an associated published research article, to which we refer readers seeking further details.

\subsection{Data Sources}
The pipeline has more than 50 input sources. They include non-profit organizations such as Crossref, preprint servers such as arXiv, academic publishers through negotiated agreements, and our own internet web crawler. Sources may provide metadata in the  JATS\footnote{\url{https://jats.nlm.nih.gov/}} format, or a variety of proprietary formats. Data may be pushed or pulled via FTP, fetched via an API, or downloaded in bulk via HTTP. Most sources are updated daily. An important additional source is human-created data. Users of the web site can claim the identity of an author in our corpus and manually curate the papers associated with that author. Our team also makes manual corrections in response to email requests from users. The first task of the pipeline is to fetch the latest data from each source and parse it into a normalized format. Sources typically provide limited information about a paper in structured form: typically the title, author names, venue, and date, often linked to a PDF file.

\begin{table}
  \centering
  \setlength{\tabcolsep}{12pt}
  \begin{tabular}{rr}
    \toprule
     \multicolumn{2}{c}{\textbf{Nodes}} \\
    \midrule
    paper & ~225M\\
    author & ~105M\\
    publication venue & ~195k\\
    \toprule
    \multicolumn{2}{c}{\textbf{Edges}} \\
    \midrule 
    citation & ~2.8B\\
    paper-author & ~650M\\
    paper-venue & 65M \\
  \bottomrule
\end{tabular}
  \caption{Approximate size of S2AG as of May 2025.}
  \label{tab:freq}
\end{table}

\subsection{PDF Content Extraction}
We augment the structured metadata by parsing the unstructured information in a paper's PDF into structured form, resulting in fine-grained information about the complete text. A critical output of our PDF content extraction is a structured bibliography from which we can construct the citation graph. Other important outputs include section headers, paragraphs, figures and tables, and inline references to figures, tables, and bibliography entries. PDFs are fundamentally a print-format specification, and extracting structured information from them is difficult and subject to errors. Nevertheless, they are the de-facto representation of a paper's official content, and we have invested heavily in our extraction technology. There are three steps to PDF content extraction: \emph{Text Extraction}, \emph{Visual Region Annotation}, and \emph{Text Span Annotation}. 

\emph{Text Extraction} is the process of converting the set of commands that indicate where characters should appear on the page into plain text, i.e., inferring word boundaries and word order. For this, we use existing open-source toolkits, including pdfalto,\footnote{\url{https://github.com/kermitt2/pdfalto}} PDFPlumber,\footnote{\url{https://github.com/jsvine/pdfplumber}} and/or PDFMiner.\footnote{\url{https://github.com/pdfminer/pdfminer.six}}

\begin{table}[t]
\small
\centering
\begin{tabular}{r|r}
\toprule
Field of Study & Count  \\
\hline
n/a  &  79M \\
Medicine  &  58M \\
Engineering  &  30M \\
Environmental Science  &  22M \\
Biology  &  21M \\
Computer Science  &  19M \\
Chemistry  &  18M \\
Physics  &  16M \\
Materials Science  &  16M \\
Psychology  &  9.9M \\
Sociology  &  8.1M \\
Political Science  &  8.0M \\
Business  &  7.9M \\
Education  &  7.5M \\
History  &  7.4M \\
Mathematics  &  6.9M \\
Economics  &  6.6M \\
Agricultural and Food Sciences  &  5.7M \\
Geography  &  4.6M \\
Geology  &  3.7M \\
Art  &  3.6M \\
Law  &  3.2M \\
Philosophy  &  3.0M \\
Linguistics  &  2.0M \\
Agricultural And Food Sciences  &  83k \\
\bottomrule
\end{tabular}
\caption{Paper record counts in S2AG for different academic fields, as of May 2025}
\label{tab:fos}
\end{table}

It is possible to extract decent-quality content using only the plain text from the Text Extraction step. For this, we have used Grobid\footnote{\url{https://github.com/kermitt2/grobid}} as well as our own extractor called  ScienceParse.\footnote{\url{https://github.com/allenai/science-parse}}$^,$\footnote{\url{https://github.com/allenai/spv2}} Our latest pipeline uses visual recognition, described below, to improve the quality of the extraction.

The \emph{Visual Region Annotation} step first generates a visual image for each page, using  poppler.\footnote{\url{https://poppler.freedesktop.org/}}. Within each page, we run an object detection model\footnote{EfficientDet~\cite{tan2020efficientdet} model trained on PubLayNet~\cite{zhong2019publaynet}} from the LayoutParser~\cite{shen2021layoutparser} library which identifies bounding boxes and labels each one with visual layout categories such as figure, table, paragraph. 
The code, model, and data are available at \href{https://github.com/Layout-Parser/layout-parser}{https://github.com/Layout-Parser/layout-parser}.


\emph{Text Span Annotation} is the process of giving semantic labels to tokens from the Text Extraction step.  For this, we use the I-VILA model trained on the S2-VL dataset, both introduced in \citet{shen2022vila}, to tag tokens with categories such as title, author name, section header, body text, figure caption, bibliography, etc. The code, model, and data are available at \href{https://github.com/allenai/VILA}{https://github.com/allenai/VILA}.

Visual Region and Text Span annotations are converted into structured data (e.g., section headers are associated with section content, figures are associated with their captions, and bibliography entries are decomposed into their constituent elements) using the PaperMage library, which makes its code, models and data available at \href{https://github.com/allenai/papermage}{https://github.com/allenai/papermage}. The final output of the PDF extraction pipeline is a structured data object suitable for encoding in JSON format. It includes core metadata such as title, authors, and abstract, as well as the full body text with detailed structural information about the text.

\begin{table*}
  \small
  \begin{center}
    \def\arraystretch{1.20}
    \begin{tabular}{rccc} 
    \toprule
    {\bf Task} & {\bf Model} & {\bf Datasets} & {\bf GitHub} \\ 
    \midrule
    
     {\begin{tabular}{@{}r@{}} Visual Region \\ Annotation\end{tabular} }  & \begin{tabular}{@{}c@{}} EfficientDet~\cite{tan2020efficientdet} via \\ LayoutParser~\cite{shen2021layoutparser} \end{tabular} 
    & \begin{tabular}{@{}c@{}} PubLayNet \\ \cite{zhong2019publaynet} \end{tabular}  &  \href{https://github.com/Layout-Parser/layout-parser}{Layout-Parser/layout-parser}   \\
    \hdashline[0.5pt/0.5pt]\noalign{\vskip 0.5ex}
    
    {\begin{tabular}{@{}r@{}} Text Span \\ Annotation\end{tabular} } & {\begin{tabular}{@{}c@{}} I-VILA \\ \cite{shen2022vila} \end{tabular} }
    & {\begin{tabular}{@{}c@{}} S2-VL \\ \cite{shen2022vila} \end{tabular} } & \href{https://github.com/allenai/VILA}{allenai/VILA} \\ 
    \hdashline[0.5pt/0.5pt]\noalign{\vskip 0.5ex}
    
    {\begin{tabular}{@{}r@{}} Paper \\ Deduplication\end{tabular} } & S2APLER & S2APLER & \href{https://github.com/allenai/S2APLER}{allenai/S2APLER} \\
    \hdashline[0.5pt/0.5pt]\noalign{\vskip 0.5ex}
    
    \begin{tabular}{@{}r@{}} Author \\ Disambiguation \end{tabular} & {\begin{tabular}{@{}c@{}} S2AND \\ \cite{Subramanian2021S2ANDAB} \end{tabular} }
    & {\begin{tabular}{@{}c@{}} S2AND \\ \cite{Subramanian2021S2ANDAB} \end{tabular} } & \href{https://github.com/allenai/S2AND}{allenai/S2AND} \\
    \hdashline[0.5pt/0.5pt]\noalign{\vskip 0.5ex}
    
    \begin{tabular}{@{}r@{}} Affiliation \\ Normalization \end{tabular} & S2AFF & S2AFF & \href{https://github.com/allenai/S2AFF}{allenai/S2AFF} \\
    \hdashline[0.5pt/0.5pt]\noalign{\vskip 0.5ex}
    
    \begin{tabular}{@{}r@{}} TLDR \\ Summarization \end{tabular} & {\begin{tabular}{@{}c@{}}  BART~\cite{lewis2019bart} with \\ CATTS~\cite{Cachola2020TLDRES} \end{tabular} }
    &  {\begin{tabular}{@{}c@{}} SciTLDR \\ \cite{Cachola2020TLDRES} \end{tabular} } & \href{https://github.com/allenai/SciTLDR}{allenai/SciTLDR} \\
    \hdashline[0.5pt/0.5pt]\noalign{\vskip 0.5ex}

    \begin{tabular}{@{}r@{}} Citation Intent \\ Classification \end{tabular} & \citet{Cohan2019StructuralSF} 
    & {\begin{tabular}{@{}c@{}} SciCite \\ \cite{Cohan2019StructuralSF} \end{tabular} } & \href{https://github.com/allenai/SciCite}{allenai/SciCite} \\
    \hdashline[0.5pt/0.5pt]\noalign{\vskip 0.5ex}

    \begin{tabular}{@{}r@{}} Field of Study \\ Classification \end{tabular} & S2FOS
    & S2FOS & \href{https://github.com/allenai/s2_fos}{allenai/s2{\textunderscore}fos} \\
    \hdashline[0.5pt/0.5pt]\noalign{\vskip 0.5ex}

    \begin{tabular}{@{}r@{}} Influential Citation  \\ Classification \end{tabular} 
    & {\begin{tabular}{@{}c@{}}  \citet{Valenzuela2015IdentifyingMC} \end{tabular} } & \href{https://allenai.org/data/meaningful-citations}{meaningful-citations} & -\\
    \hdashline[0.5pt/0.5pt]\noalign{\vskip 0.5ex}

    \begin{tabular}{@{}r@{}} Paper \\ Embedding \end{tabular} & {\begin{tabular}{@{}c@{}} SPECTER \cite{Cohan2020SPECTERDR} \\ SPECTER2 \cite{Singh2023SciRepEval} \end{tabular} }
    & {\begin{tabular}{@{}c@{}} SciDocs~\cite{Cohan2020SPECTERDR} \& \\ SciRepEval~\cite{Singh2023SciRepEval}\end{tabular} } & {\begin{tabular}{@{}c@{}} \href{https://github.com/allenai/SPECTER}{allenai/SPECTER} \\ \href{https://github.com/allenai/SciRepEval}{allenai/SciRepEval} \end{tabular} }  \\

    \bottomrule
    \end{tabular}
  \end{center}
\caption{
      Selected models and datasets used by the pipeline.}
\label{tab:models}
\end{table*}

\subsection{Knowledge Graph Construction}

The output of the PDF processing above, like the structured data in our input sources, consists of plain-string names for real world entities. Those entities include authors, the institutions with which those authors are affiliated, the venues in which the papers were published, and of course, the papers themselves. To build S2AG, we must assign an ID to each real-world entity and associate each plain-string name with the appropriate ID.

\emph{Paper Deduplication} is necessary because we process data from many independent sources. Paper titles are not unique, and can vary in how they are expressed. Authors can release updated versions of a paper with a slightly modified, or almost completely different, title. Our latest paper deduplication model is named S2APLER. It works by grouping papers into blocks using title, such that papers with similar but non-identical titles end up in the same block, then scoring the pairwise similarity of papers in each block with a trained model. The model uses string-similarity features for title, abstract, author names, venue name, etc. The synthetic training dataset uses paper data from authoritative sources that provide either a PDF or a DOI.\footnote{\url{https://www.doi.org}} Paper pairs with matching DOI/PDFs from different sources are used as positive training examples. Pairs with similar titles but non-matching DOI/PDFs are used as negative examples. We make S2APLER available at \url{https://github.com/allenai/S2APLER}.

\emph{Citation Linking} is the system for finding references to one paper in another paper's bibliography. We also use the Text Span Annotation output to associate each citation link with the text of the sentence containing the citation. Citation Linking is a very similar problem to paper deduplication, except that instead of scoring the similarity between two papers, we score the similarity between a paper and a bibliography entry produced by the PDF Extraction system. We are using fuzzy text-matching heuristics on title and authors for citation linking, but anticipate that the S2APLER model can eventually be adapted to this problem.

For \emph{Publication Venue Normalization}, we combine data from Fatcat\footnote{\url{https://fatcat.wiki/}} and MAG to build a comprehensive set of normalized venues. To match normalized venues, we index all known variant titles for the venue, including ISO-4 normalization,\footnote{\url{https://en.wikipedia.org/wiki/ISO_4}} in a direct lookup table. We apply regular-expression-based rules to the unnormalized venue strings we obtain from extracted or input sources and look for exact matches in the knowledge base.

In \emph{Author Disambiguation}, we use a system named S2AND, introduced in \citet{Subramanian2021S2ANDAB}, to assign an ID to each author mention (a name of an author appearing in a particular paper). 
S2AND operates in three stages: (1) Grouping author mentions into candidate blocks, (2) Scoring similarity between records within a block using a LightGBM model~\cite{Ke2017LightGBMAH}, and (3) Clustering mentions within a block.
The similarity scoring is a  trained on a large dataset for author disambiguation also introduced in \citet{Subramanian2021S2ANDAB}. The code, model and dataset for S2AND is available at \href{https://github.com/allenai/S2AND}{https://github.com/allenai/S2AND}.

For \emph{Author Affiliation Normalization}, we link to ROR,\footnote{\url{https://ror.org}} a registry of persistent identifiers for research organizations. Our linking model is named S2AFF.  
 It first parses unnormalized affiliation strings with a trained NER model into main institute, child institute, and address components. It then fetches the top 100 candidates from a Jaccard-overlap retrieval index and ranks them using a pairwise LightGBM model \cite{Ke2017LightGBMAH} trained using internally-gathered human annotations. We make the code and data for S2AFF available at \href{https://github.com/allenai/S2AFF}{https://github.com/allenai/S2AFF}, but the links are not currently included in our datasets.

\subsection{Semantic Features}
We now turn to describing the models that provide semantic features on top of the knowledge graph.  

\subsubsection{TLDR Summarization}
To facilitate faster understanding and decision making when scanning lists of papers, we distribute short summaries of scientific papers, or TLDRs, as introduced in \citet{Cachola2020TLDRES}. 

Generating TLDRs of scientific papers can be a challenging task that involves high source compression and requires domain-specific expertise. 
We use a BART~\cite{lewis2019bart} model trained with CATTS~\cite{Cachola2020TLDRES}, training on paper titles as a scaffolding task to overcome the problem of limited annotated training data. 
We trained this model on a combined dataset consisting of examples from SciTLDR~\cite{Cachola2020TLDRES} and a separately collected set of summaries for biomedical papers from the Semantic Scholar corpus.
The code, model, and data are available at \href{https://github.com/allenai/SciTLDR}{https://github.com/allenai/SciTLDR}.





\subsubsection{Citation Intent and Influence Classification}

Citations play a critical role in scientific papers, and understanding the intent of a citation is helpful for automated analysis of scholarly literature. 
We use the model and dataset, called SciCite, introduced in \citet{Cohan2019StructuralSF} to classify each citation into one of three categories: \emph{background information}, \emph{use of methods}, or \emph{comparing results}.
The code, model, and data for SciCite are available at \href{https://github.com/allenai/SciCite}{https://github.com/allenai/SciCite}.




We also classify whether each citation is \emph{Highly Influential}.
Based on the dataset and findings from \citet{Valenzuela2015IdentifyingMC},\footnote{\url{https://allenai.org/data/meaningful-citations}} we use a feature-based heuristic: (1) Only citations between papers with no overlapping authors are considered eligible, (2) A citation is classified as Highly Influential if it appears at least three times in a sentence in which no other papers are cited, if the citing sentence contains terms such as "build upon" "following" or "inspired by", or if the citing sentence has references to tables or figures (indicating a direct comparison with the cited work).

\subsubsection{Fields-of-Study Classification}

Prior to the discontinuation of MAG, Semantic Scholar made use of the fields-of-study classifications that MAG provided, using their level 0 taxonomy. After MAG's deprecation, we deployed our own classification model, adding Education, Law, and Linguistics to the existing MAG list. These additions were based on user feedback and comparison to other popular academic data sources such as Dimensions.

We trained our own fields-of-study classifier, named S2FOS,\footnote{\url{https://blog.allenai.org/9d2f641949e5}} using a multilabel linear SVM using character n-gram TF-IDF representations (the 300k most common character unigrams to 5-grams). For training data, we manually labeled a number of publication venues, and then propagated those labels to all papers published in their respective venues. The code, model, and data are
available at \url{https://github.com/allenai/s2_fos}





\subsubsection{Paper Embeddings}
Vector representations (embeddings) of papers can be useful in a variety of downstream applications. Our pipeline uses them for author disambiguation and recommendations, and we publish embeddings so that others may use them in their own applications. We produce embeddings using SPECTER~\cite{Cohan2020SPECTERDR}, which generates document-level embeddings from SciBERT~\cite{beltagy-etal-2019-scibert}.  SPECTER takes the paper title and abstract as input, and is trained to minimize a triplet margin loss that encourages paper pairs with a citation relationship to have more similar embeddings than those without.  Its successor, SPECTER2  \cite{Singh2023SciRepEval}, provides improved performance by training on 10x more data and introducing a new method that learns adapters \cite{houlsby2019parameter} tailored to specific task formats.  We evaluated on the SciDocs benchmark, also introduced in \citet{Cohan2020SPECTERDR}, as well as a newer benchmark SciRepEval \cite{Singh2023SciRepEval}, which increases the number and difficulty of tasks. The code, models, and data are publicly available for SPECTER \href{https://github.com/allenai/SPECTER}{https://github.com/allenai/SPECTER}, SPECTER2 \href{https://github.com/allenai/SPECTER2}{https://github.com/allenai/SPECTER2}, and SciRepEval \href{https://github.com/allenai/SciRepEval}{https://github.com/allenai/SciRepEval}.


\subsubsection{Recent Paper Recommendations}

We dynamically train recommendation models to surface relevant new papers to users. Our recommender takes a set of positively- and negatively-annotated papers, and outputs a ranked list of recommended papers. The model generates recommendations in three steps: Ranker Training, Candidate Selection, and Candidate Ranking. 

Ranker Training is an extension of Arxiv Sanity.\footnote{\url{https://arxiv-sanity-lite.com/}}$^,$\footnote{\url{https://github.com/karpathy/arxiv-sanity-lite}} From each user's positive/negative annotations, it trains two linear Support Vector Machine models: one from the TF-IDF representations of the annotated papers and one from the SPECTER embeddings.  We augment negative user annotations with randomly selected negative examples, the latter having less weight.  During Candidate Selection, we use FAISS\footnote{\url{https://github.com/facebookresearch/faiss}} to search an approximate k-nearest neighbor index of the SPECTER embeddings of $\sim$1M papers published in the last 60 days (refreshed nightly). Finally, for Candidate Ranking we select $\sim$500 papers nearest the centroid of the positively-annotated papers and rank them using the average of the two model scores.
\section{APIs and Datasets}
\label{sec:online_resources}

The outputs of our data processing pipeline and semantic models are made available through a suite of APIs and datasets described below. Because we develop and refine our models over time, the data served by the APIs may shift, or it may come from a mixture of models as we migrate from one system to its successor. Where appropriate, we will update our live documentation\footnote{\url{https://api.semanticscholar.org/api-docs}} to reflect any changes.

For unauthenticated users, we offer a low volume of API requests and samples of the datasets. For full datasets and high request volumes, we ask that users obtain an authentication key, at no charge, subject to terms of use.\footnote{\url{https://www.semanticscholar.org/product/api##Partner-Form}} To date, we have issued over 3000 authentication keys to various partners, for uses varying from student research projects to non-profit organizations to commercial products. The API served over 1.8 billion requests in 2024, with over 100 clients making more than 1k requests/day and the heaviest users making over 100k/day. The most popular types of request are paper metadata lookup by ID (60\%) and keyword search (16\%).

\begin{table}[t]
\small
\centering
\begin{tabular}{r|r}
\toprule
Field of Study & Count  \\
\hline
Medicine & 2.9M \\
Biology & 2.2M \\
Physics & 1.2M \\
Computer Science & 810k \\
Mathematics & 580k \\
Psychology & 540k \\
Chemistry & 430k \\
Materials Science & 400k \\
Environmental Science & 400k \\
Engineering & 390k \\
Agricultural And Food Sciences & 380k \\
Education & 260k \\
Business & 230k \\
Economics & 210k \\
Political Science & 150k \\
Geology & 110k \\
Art & 50k \\
Sociology & 50k \\
History & 40k \\
Linguistics & 30k \\
Philosophy & 30k \\
Law & 20k \\
Geography & 20k \\
\bottomrule
\end{tabular}
\caption{Full-text availability of papers in S2ORC for different academic fields}
\label{tab:s2orc}
\end{table}

\subsection{Graph and Search}
The Graph API\footnote{\url{https://api.semanticscholar.org/api-docs/graph}} provides the most current data from the Semantic Scholar Academic Graph, combining ID-based lookup with a variety of search endpoints Papers can be retrieved by our internal ID, or by using identifiers from arXiv, PubMed, DOI, and others. Papers can also be retrieved via their bidirectional citation relationship to other papers, or by author. For users needing to download large volumes of data, we recommend using the bulk dataset downloads described below. 

The "relevance search" endpoint is for traditional search engine queries, using keyword matching and metadata-based filters. It uses an elasticsearch index of paper titles, abstracts, and author names. Results are limited to 1000 matches, which are reranked using a trained LightGBM model. The reranking model emphasizes direct title matches and highly-cited papers with recent publication dates.

The "title search" endpoint is for locating a paper by title. If no matching paper is found, then an empty result is returned.

The "bulk search" endpoint is for retrieving large numbers of papers from broad queries, where ordering is not critical. Queries can be expressed using boolean logic, which is applied as a filter in addition to metadata-based filters. Matching papers can be sorted by citation count or publication date. The endpoint will paginate results, and can return up to 10M matching papers in total.

The "snippet search" endpoint is for retrieving matching text passages from the S2ORC dataset (see below). Titles, abstracts, and body text are loaded into a \href{https://docs.vespa.ai/}{Vespa} cluster, along with paper metadata for filtering. The index includes over 275M passages from over 12M papers across various fields of study. Each passage is limited to 480 tokens and truncated at sentence and section boundaries where possible, having an overlap of one sentence (up to 64 tokens) with the preceding and following passages. Passage text is embedded using {\tt mxbai-embed-large-v1} \citep{emb2024mxbai} with binary quantization, and placed into both a dense (approximate nearest neighbor) and sparse (keyword) index. Matching results are retrieved using the union of embedding and keyword-based matches, which are then ranked with a weighted sum of query-snippet embedding similarity and bm25 scores.

\subsection{Datasets}
\label{sec:datasets}
Download links to monthly snapshots of our knowledge graph can be obtained via our Datasets API.\footnote{\url{https://api.semanticscholar.org/api-docs/datasets}} We also publish incremental diffs between sequential releases. Each dataset is a collection of gzipped JSON files, where records in one dataset refer to records in other datasets by ID. The datasets are:

\begin{itemize}
\setlength\itemsep{-0.05em}
\item \texttt{papers}: Core metadata of papers
\item \texttt{abstracts}: Abstract text for papers, where allowed by licensing
\item \texttt{authors}: Core metadata of authors
\item \texttt{citations}: Citation links between papers, with citation context and intent and influential classifications
\item \texttt{embeddings}: SPECTER embeddings of papers
\item \texttt{paper-ids}: Mapping between different IDs used to identify a paper. Useful for tracking deduplication between releases.
\item \texttt{tldrs}: TLDRs for papers
\item \texttt{publication-venues}: Core metadata for publication venues
\item \texttt{S2ORC}: Introduced in \citet{Lo2020S2ORCTS}, S2ORC is the largest publicly-available collection of full text for open-access scientific papers. S2ORC's full text is annotated with automatically-identified structural and semantic elements of the paper: section headings, paragraphs, bibliography entries, inline citation mentions, table/figure references, etc.
Table \ref{tab:s2orc} shows the number of open-access full-text papers broken down by academic field. Since its original release as a static collection, S2ORC has grown in size and is being kept up-to-date as part of our PDF processing pipeline. For further details on S2ORC, we refer the reader to \citet{Lo2020S2ORCTS} and documentation at \href{https://github.com/allenai/s2orc}{https://github.com/allenai/s2orc}.
\end{itemize}

\subsection{Recommendations}
\label{sec:api_recommender}

The Recommendations API\footnote{\url{https://api.semanticscholar.org/api-docs/recommendations}} generates recommendations, selected from papers published within the past 60 days, based on positive/negative paper annotations. The caller provides at least one paper ID as a positive example, and any number of paper IDs as negative examples. The response is a relevance-ordered list of recently-published papers and their metadata.

\begin{table*}[tb]
  \small
  \begin{center}
    \def\arraystretch{1.20}
    \begin{tabular}{ llccr}
      \toprule
      \textbf{Resource} & \textbf{URL} & \textbf{Article Count} & \textbf{Access} & \textbf{Services}\\
      \midrule
        Aminer  & aminer.org &  321.5M & open & D$^{*}$\\
       \hdashline[0.5pt/0.5pt]\noalign{\vskip 0.5ex}
        arXiv & arxiv.org & 2M & open & D$^{**}$,F,S\\
       \hdashline[0.5pt/0.5pt]\noalign{\vskip 0.5ex}
         BASE & base-search.net & 180.5M & open & S \\
       \hdashline[0.5pt/0.5pt]\noalign{\vskip 0.5ex}
         CORE  & core.ac.uk & 207.3M & open & D$^{*}$, S \\ 
       \hdashline[0.5pt/0.5pt]\noalign{\vskip 0.5ex}
         Dimensions & app.dimensions.ai & 123.8M & subscription & D, F, M, S\\
       \hdashline[0.5pt/0.5pt]\noalign{\vskip 0.5ex}
        Google Scholar & scholar.google.com & \dan{?} &  - & - \\
       \hdashline[0.5pt/0.5pt]\noalign{\vskip 0.5ex}
        The Lens &  lens.org & 240.4M & subscription & D, M, S \\
       \hdashline[0.5pt/0.5pt]\noalign{\vskip 0.5ex}
         \dan{Meta} &  - & - & terminated \dan{3/31/22} & - \\
         \hdashline[0.5pt/0.5pt]\noalign{\vskip 0.5ex}
     \dan{Microsoft Academic} & - & - & terminated \dan{12/31/21} & - \\
       \hdashline[0.5pt/0.5pt]\noalign{\vskip 0.5ex}
      OpenAlex & openalex.org & 205.2M & open & D, F, M \\
       \hdashline[0.5pt/0.5pt]\noalign{\vskip 0.5ex}
       PubMed Central & ncbi.nlm.nih.gov/pmc/ & 7.5M & open & D$^{**}$,F,P,S\\
       \hdashline[0.5pt/0.5pt]\noalign{\vskip 0.5ex}
         ResearchGate & researchgate.net & 135.0M & - & -\\
       \hdashline[0.5pt/0.5pt]\noalign{\vskip 0.5ex}
       Scopus & scopus.com & 84.0M & subscription & F, M, S\\
       \hdashline[0.5pt/0.5pt]\noalign{\vskip 0.5ex}
      {\bf Semantic Scholar} & {\bf semanticscholar.org} & {\bf 225M}  & {\bf open} & {\bf D, \dan{F}, M, P, S, T} \\
       \hdashline[0.5pt/0.5pt]\noalign{\vskip 0.5ex}
       Web of Science Core & webofknowledge.com & 83.2M & subscription & F, M, S\\
       \bottomrule
    \end{tabular}
    \vspace{3mm}
      \begin{tabular}{r l}
         Key: &
  D=data download; 
 F=\dan{field-of-study classification; }
  M=advanced metadata; \\
  & P=semantically parsed text;
  S=\jb{title and abstract} search;
  T=natural language summarization\\
  & *=data more than a year stale; **=restricted fields of study \\
  & Article count does not include patents or datasets.
    \end{tabular}
  \end{center}
  \vspace{-5mm}
  \caption{Comparison of leading scholarly data providers}
  \label{t:compare}
  \label{tab:table1}
\end{table*}

\section{Related Work}
\label{sec:related}


Table~\ref{t:compare} summarizes major providers of scholarly data along three key dimensions: comprehensiveness, access, and services offered. Some providers, such as Google Scholar, do not offer any programmatic services at all. Others, such as MAG, have been discontinued. The major open provider of parsed content, PubMed Central, is not cross-disciplinary. Other providers require a subscription. Semantic Scholar is unique in providing a comprehensive and open knowledge base with the widest array of services.

\section{Conclusion and Future Work}
\label{sec:conclusion}

We have described the Semantic Scholar data platform, which offers code bases, data sets, and APIs covering scientific literature.  The Semantic Scholar Academic Graph (S2AG) consists of hundreds of millions of papers and billions of citation links, created by a state-of-the-art PDF extraction and knowledge graph normalization pipeline described in this paper.  The platform also offers semantic features such as summarization, vector embeddings and recommendations.  

In the future, we hope to expose selected semantic features as services, for users to apply to their own data. We hope to add richer semantic labels to our full-text annotations. We plan to add more personalized functionality, such as access to  library content and reading history. We will expand our tools for collecting human data corrections, and possibly collect automated annotations from external collaborators. Of course, we will continue to improve our existing knowledge graph construction and semantic feature models. We hope that providing these resources will enable application development and research using scholarly data to promote the advancement of science globally.

\section*{Acknowledgements}

The Semantic Scholar Open Data Platform, including S2AG and various dataset and API offerings, is the product of years of work by members of the Semantic Scholar team. The authors of this paper have all contributed directly to the creation and continued maintenance of this platform, including software and model development, data curation, evaluation, design, product management, and more.

This work was supported in part by NSF Grant CNS-2213656.

\bibliographystyle{ACM-Reference-Format}
\bibliography{main,arman,cite,dan,general,jb,kyle,linda,lucy,shannon}


\begin{thebibliography}{18}


\ifx \showCODEN    \undefined \def \showCODEN     #1{\unskip}     \fi
\ifx \showDOI      \undefined \def \showDOI       #1{#1}\fi
\ifx \showISBNx    \undefined \def \showISBNx     #1{\unskip}     \fi
\ifx \showISBNxiii \undefined \def \showISBNxiii  #1{\unskip}     \fi
\ifx \showISSN     \undefined \def \showISSN      #1{\unskip}     \fi
\ifx \showLCCN     \undefined \def \showLCCN      #1{\unskip}     \fi
\ifx \shownote     \undefined \def \shownote      #1{#1}          \fi
\ifx \showarticletitle \undefined \def \showarticletitle #1{#1}   \fi
\ifx \showURL      \undefined \def \showURL       {\relax}        \fi
\providecommand\bibfield[2]{#2}
\providecommand\bibinfo[2]{#2}
\providecommand\natexlab[1]{#1}
\providecommand\showeprint[2][]{arXiv:#2}

\bibitem[\protect\citeauthoryear{Ammar, Groeneveld, Bhagavatula, Beltagy,
  Crawford, Downey, Dunkelberger, Elgohary, Feldman, Ha, Kinney, Kohlmeier, Lo,
  Murray, Ooi, Peters, Power, Skjonsberg, Wang, Wilhelm, Yuan, van Zuylen, and
  Etzioni}{Ammar et~al\mbox{.}}{2018}]%
        {Ammar2018ConstructionOT}
\bibfield{author}{\bibinfo{person}{Waleed Ammar}, \bibinfo{person}{Dirk
  Groeneveld}, \bibinfo{person}{Chandra Bhagavatula}, \bibinfo{person}{Iz
  Beltagy}, \bibinfo{person}{Miles Crawford}, \bibinfo{person}{Doug Downey},
  \bibinfo{person}{Jason Dunkelberger}, \bibinfo{person}{Ahmed Elgohary},
  \bibinfo{person}{Sergey Feldman}, \bibinfo{person}{Vu~A. Ha},
  \bibinfo{person}{Rodney~Michael Kinney}, \bibinfo{person}{Sebastian
  Kohlmeier}, \bibinfo{person}{Kyle Lo}, \bibinfo{person}{Tyler~C. Murray},
  \bibinfo{person}{Hsu-Han Ooi}, \bibinfo{person}{Matthew~E. Peters},
  \bibinfo{person}{Joanna~L. Power}, \bibinfo{person}{Sam Skjonsberg},
  \bibinfo{person}{Lucy~Lu Wang}, \bibinfo{person}{Christopher Wilhelm},
  \bibinfo{person}{Zheng Yuan}, \bibinfo{person}{Madeleine van Zuylen}, {and}
  \bibinfo{person}{Oren Etzioni}.} \bibinfo{year}{2018}\natexlab{}.
\newblock \showarticletitle{Construction of the Literature Graph in Semantic
  Scholar}. In \bibinfo{booktitle}{\emph{NAACL}}.
\newblock
\urldef\tempurl%
\url{https://doi.org/10.18653/v1/N18-3011}
\showDOI{\tempurl}


\bibitem[\protect\citeauthoryear{Beltagy, Lo, and Cohan}{Beltagy
  et~al\mbox{.}}{2019}]%
        {beltagy-etal-2019-scibert}
\bibfield{author}{\bibinfo{person}{Iz Beltagy}, \bibinfo{person}{Kyle Lo},
  {and} \bibinfo{person}{Arman Cohan}.} \bibinfo{year}{2019}\natexlab{}.
\newblock \showarticletitle{{S}ci{BERT}: A Pretrained Language Model for
  Scientific Text}. In \bibinfo{booktitle}{\emph{Proceedings of the 2019
  Conference on Empirical Methods in Natural Language Processing and the 9th
  International Joint Conference on Natural Language Processing
  (EMNLP-IJCNLP)}}. \bibinfo{publisher}{Association for Computational
  Linguistics}, \bibinfo{address}{Hong Kong, China},
  \bibinfo{pages}{3615--3620}.
\newblock
\urldef\tempurl%
\url{https://doi.org/10.18653/v1/D19-1371}
\showDOI{\tempurl}


\bibitem[\protect\citeauthoryear{Cachola, Lo, Cohan, and Weld}{Cachola
  et~al\mbox{.}}{2020}]%
        {Cachola2020TLDRES}
\bibfield{author}{\bibinfo{person}{Isabel Cachola}, \bibinfo{person}{Kyle Lo},
  \bibinfo{person}{Arman Cohan}, {and} \bibinfo{person}{Daniel~S. Weld}.}
  \bibinfo{year}{2020}\natexlab{}.
\newblock \showarticletitle{TLDR: Extreme Summarization of Scientific
  Documents}. In \bibinfo{booktitle}{\emph{FINDINGS}}.
\newblock
\urldef\tempurl%
\url{https://doi.org/10.18653/v1/2020.findings-emnlp.428}
\showDOI{\tempurl}


\bibitem[\protect\citeauthoryear{Cohan, Ammar, van Zuylen, and Cady}{Cohan
  et~al\mbox{.}}{2019}]%
        {Cohan2019StructuralSF}
\bibfield{author}{\bibinfo{person}{Arman Cohan}, \bibinfo{person}{Waleed
  Ammar}, \bibinfo{person}{Madeleine van Zuylen}, {and} \bibinfo{person}{Field
  Cady}.} \bibinfo{year}{2019}\natexlab{}.
\newblock \showarticletitle{Structural Scaffolds for Citation Intent
  Classification in Scientific Publications}.
\newblock \bibinfo{journal}{\emph{NAACL}} (\bibinfo{year}{2019}).
\newblock


\bibitem[\protect\citeauthoryear{Cohan, Feldman, Beltagy, Downey, and
  Weld}{Cohan et~al\mbox{.}}{2020}]%
        {Cohan2020SPECTERDR}
\bibfield{author}{\bibinfo{person}{Arman Cohan}, \bibinfo{person}{Sergey
  Feldman}, \bibinfo{person}{Iz Beltagy}, \bibinfo{person}{Doug Downey}, {and}
  \bibinfo{person}{Daniel~S. Weld}.} \bibinfo{year}{2020}\natexlab{}.
\newblock \showarticletitle{SPECTER: Document-level Representation Learning
  using Citation-informed Transformers}.
\newblock \bibinfo{journal}{\emph{arXiv}}  \bibinfo{volume}{2004.07180}
  (\bibinfo{year}{2020}).
\newblock
\urldef\tempurl%
\url{https://doi.org/10.18653/v1/2020.acl-main.207}
\showDOI{\tempurl}


\bibitem[\protect\citeauthoryear{Houlsby, Giurgiu, Jastrzebski, Morrone,
  De~Laroussilhe, Gesmundo, Attariyan, and Gelly}{Houlsby
  et~al\mbox{.}}{2019}]%
        {houlsby2019parameter}
\bibfield{author}{\bibinfo{person}{Neil Houlsby}, \bibinfo{person}{Andrei
  Giurgiu}, \bibinfo{person}{Stanislaw Jastrzebski}, \bibinfo{person}{Bruna
  Morrone}, \bibinfo{person}{Quentin De~Laroussilhe}, \bibinfo{person}{Andrea
  Gesmundo}, \bibinfo{person}{Mona Attariyan}, {and} \bibinfo{person}{Sylvain
  Gelly}.} \bibinfo{year}{2019}\natexlab{}.
\newblock \showarticletitle{Parameter-Efficient Transfer Learning for {NLP}}.
  In \bibinfo{booktitle}{\emph{Proceedings of the 36th International Conference
  on Machine Learning}}.
\newblock


\bibitem[\protect\citeauthoryear{Ke, Meng, Finley, Wang, Chen, Ma, Ye, and
  Liu}{Ke et~al\mbox{.}}{2017}]%
        {Ke2017LightGBMAH}
\bibfield{author}{\bibinfo{person}{Guolin Ke}, \bibinfo{person}{Qi Meng},
  \bibinfo{person}{Thomas Finley}, \bibinfo{person}{Taifeng Wang},
  \bibinfo{person}{Wei Chen}, \bibinfo{person}{Weidong Ma},
  \bibinfo{person}{Qiwei Ye}, {and} \bibinfo{person}{Tie-Yan Liu}.}
  \bibinfo{year}{2017}\natexlab{}.
\newblock \showarticletitle{LightGBM: A Highly Efficient Gradient Boosting
  Decision Tree}. In \bibinfo{booktitle}{\emph{NIPS}}.
\newblock


\bibitem[\protect\citeauthoryear{Lee, Shakir, Koenig, and Lipp}{Lee
  et~al\mbox{.}}{2024}]%
        {emb2024mxbai}
\bibfield{author}{\bibinfo{person}{Sean Lee}, \bibinfo{person}{Aamir Shakir},
  \bibinfo{person}{Darius Koenig}, {and} \bibinfo{person}{Julius Lipp}.}
  \bibinfo{year}{2024}\natexlab{}.
\newblock \bibinfo{booktitle}{\emph{Open Source Strikes Bread - New Fluffy
  Embeddings Model}}.
\newblock
\urldef\tempurl%
\url{https://www.mixedbread.ai/blog/mxbai-embed-large-v1}
\showURL{%
\tempurl}


\bibitem[\protect\citeauthoryear{Lewis, Liu, Goyal, Ghazvininejad, Mohamed,
  Levy, Stoyanov, and Zettlemoyer}{Lewis et~al\mbox{.}}{2019}]%
        {lewis2019bart}
\bibfield{author}{\bibinfo{person}{Mike Lewis}, \bibinfo{person}{Yinhan Liu},
  \bibinfo{person}{Naman Goyal}, \bibinfo{person}{Marjan Ghazvininejad},
  \bibinfo{person}{Abdelrahman Mohamed}, \bibinfo{person}{Omer Levy},
  \bibinfo{person}{Ves Stoyanov}, {and} \bibinfo{person}{Luke Zettlemoyer}.}
  \bibinfo{year}{2019}\natexlab{}.
\newblock \showarticletitle{Bart: Denoising sequence-to-sequence pre-training
  for natural language generation, translation, and comprehension}.
\newblock \bibinfo{journal}{\emph{arXiv preprint arXiv:1910.13461}}
  (\bibinfo{year}{2019}).
\newblock


\bibitem[\protect\citeauthoryear{Lo, Wang, Neumann, Kinney, and Weld}{Lo
  et~al\mbox{.}}{2020}]%
        {Lo2020S2ORCTS}
\bibfield{author}{\bibinfo{person}{Kyle Lo}, \bibinfo{person}{Lucy~Lu Wang},
  \bibinfo{person}{Mark Neumann}, \bibinfo{person}{Rodney~Michael Kinney},
  {and} \bibinfo{person}{Daniel~S. Weld}.} \bibinfo{year}{2020}\natexlab{}.
\newblock \showarticletitle{S2ORC: The Semantic Scholar Open Research Corpus}.
  In \bibinfo{booktitle}{\emph{ACL}}.
\newblock
\urldef\tempurl%
\url{https://doi.org/10.18653/V1/2020.ACL-MAIN.447}
\showDOI{\tempurl}


\bibitem[\protect\citeauthoryear{Shen, Lo, Wang, Kuehl, Weld, and Downey}{Shen
  et~al\mbox{.}}{2022}]%
        {shen2022vila}
\bibfield{author}{\bibinfo{person}{Zejiang Shen}, \bibinfo{person}{Kyle Lo},
  \bibinfo{person}{Lucy~Lu Wang}, \bibinfo{person}{Bailey Kuehl},
  \bibinfo{person}{Daniel~S Weld}, {and} \bibinfo{person}{Doug Downey}.}
  \bibinfo{year}{2022}\natexlab{}.
\newblock \showarticletitle{VILA: Improving structured content extraction from
  scientific PDFs using visual layout groups}.
\newblock \bibinfo{journal}{\emph{Transactions of the Association for
  Computational Linguistics}}  \bibinfo{volume}{10} (\bibinfo{year}{2022}),
  \bibinfo{pages}{376--392}.
\newblock


\bibitem[\protect\citeauthoryear{Shen, Zhang, Dell, Lee, Carlson, and Li}{Shen
  et~al\mbox{.}}{2021}]%
        {shen2021layoutparser}
\bibfield{author}{\bibinfo{person}{Zejiang Shen}, \bibinfo{person}{Ruochen
  Zhang}, \bibinfo{person}{Melissa Dell}, \bibinfo{person}{Benjamin
  Charles~Germain Lee}, \bibinfo{person}{Jacob Carlson}, {and}
  \bibinfo{person}{Weining Li}.} \bibinfo{year}{2021}\natexlab{}.
\newblock \showarticletitle{LayoutParser: A unified toolkit for deep learning
  based document image analysis}. In \bibinfo{booktitle}{\emph{International
  Conference on Document Analysis and Recognition}}. Springer,
  \bibinfo{pages}{131--146}.
\newblock


\bibitem[\protect\citeauthoryear{Singh, D{'}Arcy, Cohan, Downey, and
  Feldman}{Singh et~al\mbox{.}}{2023}]%
        {Singh2023SciRepEval}
\bibfield{author}{\bibinfo{person}{Amanpreet Singh}, \bibinfo{person}{Mike
  D{'}Arcy}, \bibinfo{person}{Arman Cohan}, \bibinfo{person}{Doug Downey},
  {and} \bibinfo{person}{Sergey Feldman}.} \bibinfo{year}{2023}\natexlab{}.
\newblock \showarticletitle{{S}ci{R}ep{E}val: A Multi-Format Benchmark for
  Scientific Document Representations}. In
  \bibinfo{booktitle}{\emph{Proceedings of the 2023 Conference on Empirical
  Methods in Natural Language Processing}},
  \bibfield{editor}{\bibinfo{person}{Houda Bouamor}, \bibinfo{person}{Juan
  Pino}, {and} \bibinfo{person}{Kalika Bali}} (Eds.).
  \bibinfo{publisher}{Association for Computational Linguistics},
  \bibinfo{address}{Singapore}, \bibinfo{pages}{5548--5566}.
\newblock
\urldef\tempurl%
\url{https://doi.org/10.18653/v1/2023.emnlp-main.338}
\showDOI{\tempurl}


\bibitem[\protect\citeauthoryear{Sinha, Shen, Song, Ma, Eide, Hsu, and
  Wang}{Sinha et~al\mbox{.}}{2015}]%
        {Sinha2015AnOO}
\bibfield{author}{\bibinfo{person}{Arnab Sinha}, \bibinfo{person}{Zhihong
  Shen}, \bibinfo{person}{Yang Song}, \bibinfo{person}{Hao Ma},
  \bibinfo{person}{Darrin Eide}, \bibinfo{person}{Bo-June~Paul Hsu}, {and}
  \bibinfo{person}{Kuansan Wang}.} \bibinfo{year}{2015}\natexlab{}.
\newblock \showarticletitle{An Overview of Microsoft Academic Service (MAS) and
  Applications}.
\newblock \bibinfo{journal}{\emph{Proceedings of the 24th International
  Conference on World Wide Web}} (\bibinfo{year}{2015}).
\newblock
\urldef\tempurl%
\url{https://doi.org/10.1145/2740908.2742839}
\showDOI{\tempurl}


\bibitem[\protect\citeauthoryear{Subramanian, King, Downey, and
  Feldman}{Subramanian et~al\mbox{.}}{2021}]%
        {Subramanian2021S2ANDAB}
\bibfield{author}{\bibinfo{person}{Shivashankar Subramanian},
  \bibinfo{person}{Daniel King}, \bibinfo{person}{Doug Downey}, {and}
  \bibinfo{person}{Sergey Feldman}.} \bibinfo{year}{2021}\natexlab{}.
\newblock \showarticletitle{S2AND: A Benchmark and Evaluation System for Author
  Name Disambiguation}.
\newblock \bibinfo{journal}{\emph{ArXiv}}  \bibinfo{volume}{abs/2103.07534}
  (\bibinfo{year}{2021}).
\newblock


\bibitem[\protect\citeauthoryear{Tan, Pang, and Le}{Tan et~al\mbox{.}}{2020}]%
        {tan2020efficientdet}
\bibfield{author}{\bibinfo{person}{Mingxing Tan}, \bibinfo{person}{Ruoming
  Pang}, {and} \bibinfo{person}{Quoc~V Le}.} \bibinfo{year}{2020}\natexlab{}.
\newblock \showarticletitle{Efficientdet: Scalable and efficient object
  detection}. In \bibinfo{booktitle}{\emph{Proceedings of the IEEE/CVF
  conference on computer vision and pattern recognition}}.
  \bibinfo{pages}{10781--10790}.
\newblock


\bibitem[\protect\citeauthoryear{Valenzuela, Ha, and Etzioni}{Valenzuela
  et~al\mbox{.}}{2015}]%
        {Valenzuela2015IdentifyingMC}
\bibfield{author}{\bibinfo{person}{Marco Valenzuela}, \bibinfo{person}{Vu~A.
  Ha}, {and} \bibinfo{person}{Oren Etzioni}.} \bibinfo{year}{2015}\natexlab{}.
\newblock \showarticletitle{Identifying Meaningful Citations}. In
  \bibinfo{booktitle}{\emph{AAAI Workshop: Scholarly Big Data}}.
\newblock


\bibitem[\protect\citeauthoryear{Zhong, Tang, and Yepes}{Zhong
  et~al\mbox{.}}{2019}]%
        {zhong2019publaynet}
\bibfield{author}{\bibinfo{person}{Xu Zhong}, \bibinfo{person}{Jianbin Tang},
  {and} \bibinfo{person}{Antonio~Jimeno Yepes}.}
  \bibinfo{year}{2019}\natexlab{}.
\newblock \showarticletitle{PubLayNet: largest dataset ever for document layout
  analysis}. In \bibinfo{booktitle}{\emph{2019 International Conference on
  Document Analysis and Recognition (ICDAR)}}. IEEE,
  \bibinfo{pages}{1015--1022}.
\newblock
\showISSN{1520-5363}
\urldef\tempurl%
\url{https://doi.org/10.1109/ICDAR.2019.00166}
\showDOI{\tempurl}


\end{thebibliography}

\end{document}